\documentclass[12pt,a4paper]{article}
\pdfoutput=1
\usepackage{color}
\usepackage{amssymb,amsmath,bm,bbold}
\usepackage{epsf}
\usepackage{epsfig}
\usepackage{relsize}
\usepackage[dvipsnames]{xcolor}
\usepackage[linktoc=page,bookmarks=false,colorlinks=false,linkbordercolor=RoyalBlue,citebordercolor=ForestGreen,urlbordercolor=CornflowerBlue]{hyperref}
\usepackage{latexsym,mathrsfs,dsfont}
\usepackage[normalem]{ulem} 
\usepackage[compress]{cite}
\usepackage{graphicx}
\usepackage{url}
\usepackage{booktabs}
\usepackage{float}
\usepackage{multirow}
\usepackage{changepage}
\usepackage[hypcap]{caption, subcaption}

\usepackage[titles]{tocloft}

\usepackage{tabularx,colortbl}

\usepackage{fancyhdr}
\advance \headheight by 3.0truept       

\allowdisplaybreaks[1]

\definecolor{red}{cmyk}{0,1,1,0.4}
\definecolor{darkgreen}{rgb}{0.0,0.6,0.0}
\definecolor{cDarkGrey}{RGB}{91,91,91}
\definecolor{cGrey}{RGB}{245,243,238}
\definecolor{cBlue}{RGB}{0,110,191}
\definecolor{cLightBlue}{RGB}{214,237,252}
\definecolor{cRed}{RGB}{196,0,100}
\definecolor{cLightRed}{RGB}{254,222,237}
\definecolor{cGreen}{RGB}{0,166,80}
\definecolor{cLightGreen}{RGB}{254,222,237}
\definecolor{cOrange}{RGB}{221,74,44}
\definecolor{cLightOrange}{RGB}{255,215,210}
\definecolor{cPurple}{RGB}{93,35,125}
\definecolor{cLightPurple}{RGB}{241,230,252}
\definecolor{cYellow}{RGB}{252,191,10}
\definecolor{cISSRBlue}{RGB}{0,111,174}
\definecolor{cISSRGrey}{RGB}{167,169,172}

\newcommand{\beq}{\begin{equation}}
\newcommand{\eeq}{\end{equation}}
\newcommand{\be}{\begin{equation}}
\newcommand{\ee}{\end{equation}}
\newcommand{\bi}{\begin{itemize}}
\newcommand{\ei}{\end{itemize}}
\newcommand{\ba}{\begin{array}}
\newcommand{\ea}{\end{array}}
\newcommand{\beqa}{\begin{eqnarray}}
\newcommand{\eeqa}{\end{eqnarray}}
\newcommand{\bea}{\begin{eqnarray}}
\newcommand{\eea}{\end{eqnarray}}
\newcommand{\beqn}{\begin{eqnarray}}
\newcommand{\eeqn}{\end{eqnarray}}

\newcounter{TODO}

\newcommand{\vcb}{|V_{cb}|}

\newcommand{\vub}{|V_{ub}|}

\def\kpn{K^+\rightarrow\pi^+\nu\bar\nu}
\def\klpn{K_{L}\rightarrow\pi^0\nu\bar\nu}

\begin{document}

\begin{flushleft}
\end{flushleft}

\vspace{-14mm}
\begin{flushright}
  AJB-22-8
\end{flushright}

\medskip

\begin{center}
{\Large\bf\boldmath
   On the Standard Model Predictions for\\ Rare
    K and B Decay Branching Ratios: 2022
}
\\[1.0cm]
{\bf
    Andrzej~J.~Buras
}\\[0.3cm]

{\small
TUM Institute for Advanced Study,
    Lichtenbergstr. 2a, D-85747 Garching, Germany \\[0.2cm]
Physik Department, TU M\"unchen, James-Franck-Stra{\ss}e, D-85748 Garching, Germany
}
\end{center}

\vskip 0.5cm

\begin{abstract}
  \noindent
  In this  decade one expects a very significant progress
  in measuring the branching ratios for several rare $K$ and $B$ decays,
  in particular for the decays $K^+\to\pi^+\nu\bar\nu$, $K_L\to\pi^0\nu\bar\nu$, $B_s\to\mu^+\mu^-$ and   $B_d\to\mu^+\mu^-$. 
On the theory   side a very significant progress on calculating these branching ratios has been achieved in the last thirty years culminating recently in rather precise SM predictions   for them. It is then {\em unfortunate} that   some papers still cite the results for $K^+\to\pi^+\nu\bar\nu$ and
  $K_L\to\pi^0\nu\bar\nu$   presented by us in 2015. They are clearly out of date.  Similar  comments apply to predictions for $B_{s,d}\to\mu^+\mu^-$.
In this note I want to stress again that, in view of the tensions between various determinations of $\vcb$ in tree-level decays, presently, the
only trustable SM predictions for the branching ratios in question
  can be obtained by eliminating their dependence on the CKM parameters with the help
  of $|\varepsilon_K|$, $\Delta M_s$, $\Delta M_d$ and $S_{\psi K_S}$,
  evaluated in the SM, and setting their values to the experimental ones.
 This is supported by the fact that  presently  NP is not required
  to describe simultaneously   the very precise data on $|\varepsilon_K|$, $\Delta M_s$,  $\Delta M_d$ and   $S_{\psi K_S}$.   This strategy for obtaining true SM predictions for rare decay branching ratios   is moreover  not polluted   by hadronic uncertainies and observed anomalies in semi-leptonic decays   used often in global analyses as stressed recently in a longer paper by the present author
\cite{Buras:2022qip}.

  \end{abstract}

\thispagestyle{empty}
\newpage
\setcounter{page}{1}

%
%
%
\section{Introduction}
In this decade and the next decade one expects a very significant progress
  in measuring the branching ratios for several rare $K$ and $B$ decays,
  in particular for the decays $\kpn$, $\klpn$, $B_s\to\mu^+\mu^-$ and
  $B_d\to\mu^+\mu^-$ \cite{Cerri:2018ypt,Bediaga:2018lhg,NA62:2022nah}.  On the theory
  side a very significant progress on these branching ratios has been achieved in the last thirty years\footnote{We refrain from listing these references as they all can be found in the papers cited here. In particular in my book \cite{Buras:2020xsm}.} culminating recently in rather precise SM predictions
  on the four branching ratios in question \cite{Buras:2021nns,Buras:2022wpw}.
 It is then unfortunate that
 some papers and conference presentations still cite the results for $K^+\to\pi^+\nu\bar\nu$ and
  $K_L\to\pi^0\nu\bar\nu$   presented by us in 2015 \cite{Buras:2015qea}.
While this paper contains several useful expressions that could play an important  role when the data on these decays  improves, the final results for the branching ratios  presented there are clearly out of date. This is also the case of
more recent predictions of Brod, Gorbahn and Stamou \cite{Brod:2021hsj}.
Similar comments apply to predictions for $B_{s,d}\to\mu^+\mu^-$ quoted recently in the literature and at conferences.
In these analyses either specific assumptions on the values of $\vcb$
and $\vub$ from tree-level decays have been used or as in the case
of \cite{Brod:2021hsj} the results of an out-of-date global CKM fit have been
inserted in the otherwise correct SM formulae. The latter authors will soon update their result using most recent global fit of PDG \cite{Workman:2022ynf}.

In this updated note, that can be considered as an overture to the very recent paper \cite{Buras:2022qip}, I want to stress again, as done already in \cite{Buras:2021nns,Buras:2022wpw,Aebischer:2022vky,Buras:2022nrb},
that  in view of the tensions between exclusive and inclusive determinations of $\vcb$ and $\vub$ \cite{Bordone:2021oof,Aoki:2021kgd}, presently the
  only trustable SM predictions for the branching ratios in question
   can be obtained by eliminating their dependence on CKM
   parameters with the help of the $\Delta F=2$ observables
\be\label{loop}
|\varepsilon_K|,\qquad \Delta M_s,\qquad \Delta M_d, \qquad S_{\psi K_S}\,,
\ee
   evaluated in the SM. They are very well measured and the theory is under
   good control.    
  Moreover, as demonstrated in \cite{Buras:2022wpw}, with the most recent
  values of hadronic matrix elements from HPQCD collaboration \cite{Dowdall:2019bea}, NP is not required
  to describe these four observables simultaneously\footnote{In fact 
   as stressed recently
      by Gino Isidori in a talk at the third Anomaly Workshop in Durham,
      there are presently no signs of NP in $\Delta F=2$ processes.}
This strategy for obtaining true SM predictions for rare decay branching ratios   is moreover  not polluted   by hadronic uncertainies and observed anomalies in semi-leptonic decays   used often in global analyses.

      Let me make also clear that I am not against new citations to my 2015
      paper  \cite{Buras:2015qea} because
      there are several useful results in that paper, but to quote
      our 2015 results as best results in 2022, is equivalent to stating
      that theorists made no progress in the last seven years. It is like
      referring in 2022 to the Brookhaven experiment for the best measurement of
      $\kpn$ branching ratio, instead to the last
      impressive result from the
       NA62 collaboration \cite{NA62:2021zjw} 
\be\label{EXP19}
\mathcal{B}(\kpn)_\text{exp}=(10.6^{+4.0}_{-3.5}\pm 0.9)\times 10^{-11}\,.
\ee

  We will next exhibit these statements in numbers.
  
\section{The Past}
There are two estimates of $\kpn$ and $\klpn$ SM branching ratios presented by us in 2015  \cite{Buras:2015qea}
\be\label{SM1}
\mathcal{B}(\kpn)_\text{SM}= (8.4\pm 1.0)\times 10^{-11}\,, \qquad
\mathcal{B}(\klpn)_\text{SM}=(3.4\pm0.6)\times 10^{-11},
\ee
usually cited by the experimentalists, and
\be\label{SM2}
\mathcal{B}(\kpn)_\text{SM}= (9.11\pm 0.72)\times 10^{-11}\,, \qquad
\mathcal{B}(\klpn)_\text{SM}=(3.00\pm0.30)\times 10^{-11}\,.
\ee

The result in (\ref{SM1}) is based on the 2015 average of inclusive
and exclusive values of $\vcb$ and $\vub$
\be
\vcb_\text{avg}=(40.7\pm 1.4)\times 10^{-3}, \qquad
\vub_\text{avg}=(3.88\pm 0.29)\times 10^{-3}.
\ee
The result in (\ref{SM2}) is based on the 2015 determination
of $\vcb$ and $\vub$ from $\Delta F=2$ observables in (\ref{loop})
that implied
\be\label{loopCKM}
\vcb_\text{loop}=(42.4\pm 1.2)\times 10^{-3}, \qquad
\vub_\text{loop}=(3.61\pm 0.14)\times 10^{-3}.
\ee

Finally the authors of 
\cite{Brod:2021hsj} found in 2019
\be\label{Brod}
\mathcal{B}(\kpn)_\text{SM}= (7.7\pm0.6)\times 10^{-11}\,,\qquad
\mathcal{B}(\klpn)_\text{SM}=(2.6\pm0.3)\times 10^{-11}\,.
\ee

The three results in (\ref{SM1}), (\ref{SM2}) and (\ref{Brod})
are out of date for the following different reasons
\begin{itemize}
\item Result in (\ref{SM1}) uses some averages of inclusive and exclusive
  values of $\vcb$ and $\vub$ that  are
  hardly consistent with each other. As demonstrated in \cite{Buras:2021nns} the standard route to SM predictions
for $\kpn$ and $\klpn$ used in \cite{Buras:2015qea} has to be avoided
because of the known tensions in the determinations of $\vcb$ from tree-level decays.
  \item
    Result in (\ref{SM2}) uses 2015 hadronic matrix elements from Lattice
    QCD relevant for $\Delta M_s$ and $\Delta M_d$ that changed significantly
    since then.
  \item
    The authors of \cite{Brod:2021hsj} made in an earlier paper
     \cite{Brod:2019rzc} significant progress in
     the reduction of theoretical uncertainties in $\varepsilon_K$.
     Unfortunately, in predicting the branching ratios for $\kpn$ and
     $\klpn$ they used  the values of the CKM parameters
     adopted by PDG in 2020 \cite{Zyla:2020zbs}. These values are out of date
     because in obtaining them the 2016 values of hadronic matrix
     elements relevant for $\Delta M_d$ and $\Delta M_s$ have been used\footnote{Private communication from one of the CKMfitters.}.
     As demonstrated in \cite{Blanke:2016bhf,Blanke:2018cya} and recently
     in \cite{Buras:2022wpw} there are significant inconsistencies
     in the SM description of $\Delta M_d$ and $\varepsilon_K$ if the hadronic
     input of 2016 is used.
     \end{itemize}

In the case of $B_{s,d}\to\mu^+\mu^-$ branching ratios the
common values quoted in the literature and conference presentations are the ones from
Beneke, Bobeth and Szafron \cite{Beneke:2017vpq,Beneke:2019slt}
\be\label{Beneke}
\overline{\mathcal{B}}(B_{s}\to\mu^+\mu^-)_{\rm SM} = (3.66\pm0.14)\times 10^{-9},\quad \mathcal{B}(B_{d}\to\mu^+\mu^-)_{\rm SM} = (1.03\pm 0.05)\ \times 10^{-10}.
\ee
This is an important last detailed direct calculation of these branching ratios
that further reduced theoretical uncertainties, but in reality subject to larger $\vcb$ uncertainties than quoted above
as pointed out in \cite{Bobeth:2021cxm} with the participation of one of the authors of  \cite{Beneke:2017vpq,Beneke:2019slt}. To circumvent this difficulty
the 2003 proposal of the present author \cite{Buras:2003td}, see below, has been used in \cite{Bobeth:2021cxm} and also in \cite{Buras:2021nns,Buras:2022wpw}.

\section{2022}
For completeness let me finish this note by recalling the basis formulae
which were used in  \cite{Buras:2021nns,Buras:2022wpw} to obtain
the most accurate predictions for the four branching ratios in question to date.

These are the following four $\vcb$-independent ratios that are valid {\em only} within the SM:
\be\label{R11}
  \boxed{R_{11}(\beta,\gamma)=\frac{\mathcal{B}(\kpn)}{|\varepsilon_K|^{0.82}}=(1.31\pm0.05)\times 10^{-8}{\left(\frac{\sin\gamma}{\sin 67^\circ}\right)^{0.015}\left(\frac{\sin 22.2^\circ}{\sin \beta}\right)^{0.71},  }            }
  \ee
  \be\label{R12a}
\boxed{R_{12}(\beta,\gamma)=\frac{\mathcal{B}(\klpn)}{|\varepsilon_K|^{1.18}}=(3.87\pm0.06)\times 10^{-8}
    {\left(\frac{\sin\gamma}{\sin 67^\circ}\right)^{0.03}\left(\frac{\sin\beta}{\sin 22.2^\circ}\right)^{0.9{8}},}}
  \ee
  \be\label{CMFV61}
\boxed{R_q=\frac{\mathcal{B}(B_q\to\mu^+\mu^-)}{\Delta M_q}= 4.291\times 10^{-10}\ \frac{\tau_{B_q}}{\hat B_q}\frac{(Y_0(x_t))^2}{S_0(x_t)},\qquad q=d,s\,.}
\ee
The notation is known to any flavour practitioner.  The first two have been found in \cite{Buras:2021nns}, the last two already in 2003 in \cite{Buras:2003td}. 
Note that
the only relevant CKM parameter in these $\vcb$-independent ratios
 is the UT angle $\beta$ and this is the reason why we need the mixing induced CP-asymmetry $S_{\psi K_S}$ to obtain predictions for $\kpn$ and $\klpn$.
 While $\gamma$ also enters these expressions, its impact on final results is practically irrelevant. Changing $\gamma$ in this expression from $67^\circ$ to
 $64.6(16)^\circ$ quoted in (\ref{CKMoutput}) has practically no impact on the
 four ratios in question.  This is still another advantage of this strategy
 over global fits in addition to the independence of $\vcb$.

Setting the values of the four $\Delta F=2$ observables in (\ref{loop})
to their experimental values and including all experimental and theoretical uncertainties one finds
\cite{Buras:2021nns}
\be\label{BV}
\boxed{\mathcal{B}(\kpn)_\text{SM}= {(8.60\pm 0.42)}\times 10^{-11}\,,\quad
\mathcal{B}(\klpn)_\text{SM}={(2.94\pm 0.15)}\times 10^{-11}\,,}
\ee
and \cite{Buras:2022wpw}
\be\label{LHCbTH}
\boxed{\overline{\mathcal{B}}(B_{s}\to\mu^+\mu^-)_{\rm SM} = (3.78^{+ 0.15}_{-0.10})\times 10^{-9},\quad \mathcal{B}(B_{d}\to\mu^+\mu^-)_{\rm SM} = (1.02^{+ 0.05}_{-0.03})\ \times 10^{-10}.}
\ee
Note that relative to the usually quoted values by experimentalists in
(\ref{SM1}) the central values for $\kpn$ and $\klpn$ did not change by much
but the uncertainties decreased by factors $2.5$ and $4.0$, respectively.
Therefore, referring to (\ref{SM1}) as the present best values instead of
(\ref{BV}) totally misreprents the situation and should be abandoned as
already emphasized in  \cite{Buras:2021nns,Buras:2022wpw,Aebischer:2022vky,Buras:2022nrb}. The values in (\ref{LHCbTH}) are rather close to those in
(\ref{Beneke}), although the value for $B_s\to\mu^+\mu^-$ is by one $\sigma$
larger.

Indeed, the most interesting at present is the SM prediction for the $B_{s}\to\mu^+\mu^-$
branching ratio that exibits a $2.7\sigma$ anomaly when confronted with its experimental
value \cite{LHCb:2021awg,CMS:2020rox,ATLAS:2020acx}
\be\label{LHCbEXP1}
\overline{\mathcal{B}}(B_{s}\to\mu^+\mu^-)_{\rm EXP} = 2.86(33)\times 10^{-9}\,.
\ee
In terms of the ratio $R_s$ this anomaly is given by
\be\label{CMFV6}
\left[\frac{\mathcal{B}(B_s\to\mu^+\mu^-)}{\Delta M_s}\right]_{\text{SM}}=2.13(7)\times 10^{-10}\,\text{ps},\quad\left[\frac{\mathcal{B}(B_s\to\mu^+\mu^-)}{\Delta M_s}\right]_{\text{EXP}}=1.61(18)\times 10^{-10}\,\text{ps}\,.
\ee
Unfortunately the recent messages from CMS are expected to decrease or even
remove this anomaly. We are looking forward to the average from CMS, LHCb
and ATLAS.

The support for this procedure comes from the  analysis in \cite{Buras:2022wpw}. 
It turns out  that the simultaneous description of the data for $\Delta M_d$, $\Delta M_s$,  $\varepsilon_K$ and $S_{\psi K_S}$ can be made without any participation of NP. Indeed, as seen in the first $\vcb-\gamma$ plot in Fig.~8 of that paper, the SM predictions for $\varepsilon_K$, $\Delta M_d$,  $\Delta M_s$
and $S_{\psi K_S}$ turn out to be remarkably 
consistent with each other and with the data for the following values
of the CKM parameters \cite{Buras:2022wpw}
\be\label{CKMoutput}
\vcb=42.6(4)\times 10^{-3}, \quad 
\gamma=64.6(16)^\circ, \quad \beta=22.2(7)^\circ, \quad \vub=3.72(11)\times 10^{-3}\,.
\ee
As emphasized in \cite{Buras:2022wpw} this agreement
is only found using
the hadronic matrix elements with $2+1+1$ flavours from the lattice HPQCD collaboration \cite{Dowdall:2019bea}\footnote{Similar results for $\Delta M_d$ and $\Delta M_s$ hadronic
    matrix elements have been obtained within the HQET sum rules in
    \cite{Kirk:2017juj} and \cite{King:2019lal}, respectively.}
.  These values are consistent
with the inclusive determination of $\vcb$ in \cite{Bordone:2021oof}  and the exclusive ones of $\vub$ from FLAG \cite{Aoki:2021kgd}.

Please note that our strategy in \cite{Buras:2021nns}, similarly to the second 
2015 strategy in \cite{Buras:2015qea}, used as constraints the experimental values for $\Delta F=2$
observables in (\ref{loop}). However, the elimination of $\vcb$
with the help of the ratios (\ref{R11}) and  (\ref{R12a}), not done in the 2015
paper 
\cite{Buras:2015qea}, simplified the subsequent numerical analysis significantly. 

The points made in this note are discussed in greater detail in
\cite{Buras:2022qip}, where  using this strategy we obtain SM predictions for 26 branching   ratios for rare semileptonic and leptonic $K$ and $B$ decays with the $\mu^+\mu^-$ pair   or the $\nu\bar\nu$ pair in the final state.
 Most interesting   turn out to be the anomalies in the low $q^2$ bin in
 $B^+\to K^+\mu^+\mu^-$ ($5.1\sigma$) and $B_s\to \phi\mu^+\mu^-$ ($4.8\sigma$).

 Let me also emphasize  that in any global fit of CKM parameters the SM expressions
 for the four observables in (\ref{loop}) are set to their experimental
 values like in our strategy. But in usual global fits other observables are
 included which we intentionally leave out  in our strategy  to avoid the tension in  $\vcb$ and $\vub$ 
 determinations from tree-level decays and the danger of NP infection of the resulting CKM parameters and consequently of SM predictions for rare branching ratios. Moreover, the pollution by hadronic uncertainties in other decays used in the fit
 that are larger than in many rare decay branching ratios considered can be avoided in this manner.
 We have elaborated on this in \cite{Buras:2022qip}. 

 Finally, let me stress that this is a novel route to SM predictions for rare decay branching ratios that has
 not been explored by anybody to date. Only time will show whether this new
 strategy will be more successful to find NP than the usual global fits.

{\bf Acknowledgements} Many thanks to Elena Venturini for the most enjoyable collaboration. Financial support from the Excellence Cluster ORIGINS,
funded by the Deutsche Forschungsgemeinschaft (DFG, German Research
Foundation), 
Excellence Strategy, EXC-2094, 390783311 is acknowledged.
\renewcommand{\refname}{R\lowercase{eferences}}

\addcontentsline{toc}{section}{References}

\bibliographystyle{JHEP}

\small

\bibliography{Bookallrefs}

\end{document}